\documentclass{emulateapj}
\usepackage{apjfonts}

\newcommand{\PSR}{PSR~B1259$-$63}

\newcommand{\RXP}{2RXP J130159.6$-$635806}
\newcommand{\sRXP}{2RXP~J1301}
\newcommand{\pair}{$e^{\pm}$}

\slugcomment{Accepted for publication in ApJ}

\shorttitle{Suzaku Observations of PSR B1259$-$63}
\shortauthors{Uchiyama et al.}

\begin{document}


\title{Suzaku Observations of PSR B1259$-$63: 
A New Manifestation of Relativistic Pulsar Wind}


\author{Yasunobu Uchiyama,\altaffilmark{1,2} Takaaki Tanaka,\altaffilmark{1}
Tadayuki Takahashi,\altaffilmark{3,4} Koji Mori,\altaffilmark{5}
and Kazuhiro Nakazawa\altaffilmark{4}}
\email{uchiyama@slac.stanford.edu}
\altaffiltext{1}{SLAC National Accelerator Laboratory, 2575 Sand Hill Road M/S 29, Menlo Park, CA 94025, USA.}
\altaffiltext{2}{Panofsky Fellow.}
\altaffiltext{3}{Institute of Space and Astronautical Science/JAXA, Sagamihara, Kanagawa 229-8510, Japan.}
\altaffiltext{4}{Department of Physics, University of Tokyo, 7-3-1 Hongo, Bunkyoku, 
Tokyo 113-0033, Japan.}
\altaffiltext{5}{Department of Applied Physics,  University of Miyazaki, 
1-1 Gakuen Kibana-dai Nishi, Miyazaki 889-2192, Japan.}

\begin{abstract}
We observed  PSR B1259$-$63, a young non-accreting pulsar orbiting around a Be star SS 2883, 
eight times with the \emph{Suzaku} satellite from July to September 2007, 
to characterize the X-ray emission arising from the interaction between 
a pulsar relativistic wind and Be star outflows. 
The X-ray spectra showed
a featureless continuum in 0.6--10 keV, modeled by 
a power law with a wide range of photon index 1.3--1.8.
When combined with the 
\emph{Suzaku} PIN detector which allowed spectral analysis 
in the hard 15--50 keV band,  X-ray spectra do show a break 
at $\sim 5$ keV in a certain epoch.
Regarding the \PSR\ system as a  compactified pulsar wind nebula, 
in which \pair\ pairs are assumed to be accelerated at 
 the inner shock front of the pulsar wind, 
we attribute the X-ray spectral break to the low-energy cutoff 
of the synchrotron radiation associated with 
 the Lorentz factor of the relativistic pulsar wind
$\gamma_1 \sim  4\times 10^5$.
Our result indicates that 
Comptonization of stellar photons by the unshocked pulsar wind will be 
accessible (or tightly constrained) by 
observations with the \emph{Fermi} Gamma-ray Space Telescope 
during the next periastron passage.
The \PSR\ system  allows us to probe 
the fundamental properties of the pulsar wind by a direct means,  
being complementary 
to the study of large-scale pulsar wind nebulae. 
\end{abstract}


\section{Introduction}

PSR B1259$-$63 is a young radio pulsar (spin period 48 ms) orbiting 
a B2e star SS 2883 in a highly eccentric 3.4 yr orbit, 
having spindown power of 
$\dot{E}_{\rm p} \simeq 8\times 10^{35}\ \rm erg\ s^{-1}$
\citep{Johnston92,Johnston94,Manchester95}.
The distance from the Earth is estimated as $\sim 1.5$ kpc \citep{Johnston94}.
\PSR\ was 
the first example of a radio pulsar forming a binary with a non-degenerate 
companion \citep{Johnston92}.
The \PSR\ system is an important astrophysical laboratory 
for the study of the pulsar wind interacting with a stellar wind, 
and the nonthermal radiation arising from the shocked pulsar wind. 
Apart from \PSR, the only other example known  in our Galaxy 
is the PSR~J1740$-$3052 binary system \citep{Stairs01}, which has,  
however,  much lower spindown power. 

Be stars are characterized by strong mass outflows through 
the formation of a slow and dense equatorial disk \citep{Waters88}.
Because \PSR\ has a highly eccentric 
orbit and the equatorial disk is likely inclined with respect to the orbital plane, 
the pulsar  is expected to interact with the disk of the Be star near periastron, by 
crossing the disk twice before and after the periastron passage 
\citep[e.g.,][]{Melatos95}.
The signatures of disk crossings can be seen in 
the radio observations, such as pulsar eclipse, 
the light curve of transient unpulsed emission, and
the change of dispersion measure \citep{Johnston96,Johnston05}.
The disk crossings also play important roles in producing the X-ray 
and gamma-ray emission \citep{Chern06}.

Enhanced high-energy emission around periastron 
which is thought to arise from 
the interactions between the pulsar wind and the stellar wind/disk, 
has been observed in X-ray 
\citep{Kaspi95,Hirayama96,Hirayama99,Chern06}, 
soft $\gamma$-ray 
\citep{Grove95,Shaw04},
and very high energy gamma-rays \citep{HESS05}.
The basic properties of the X-ray emission have been 
revealed by the series of \emph{ASCA} observations 
\citep{Kaspi95,Hirayama96,Hirayama99}, 
which are summarized as follows:
(1) The X-ray luminosity is modest around periastron 
with $L_x \sim 10^{34}\ \rm erg\ s^{-1}$, corresponding to about 
one percent of the spindown power.
 The intensity within a day is 
constant, while it varies by a factor of $\sim 2$ 
during the periastron passage. At around apastron, 
the luminosity decreased by one order of magnitude;
(2) The X-ray spectrum is characterized by a power law of photon index 
$\Gamma \simeq 1.6\mbox{--}2.0$ without any detectable line emission;
(3) No significant X-ray pulsed emission at the pulsar spin period has been 
found.  The upper limit (at a 90\% level) on the pulse component 
is $\sim 10\%$ of the total X-ray flux.
Observations during the periastron passage in 2004 
with \emph{XMM-Newton} have revealed more detailed 
 behavior of X-ray emission \citep{Chern06}. 
The X-ray brightening prior to periastron was roughly coincident with the 
 brightening of unpulsed radio emission, which can be identified with the 
 entrance to the Be star equatorial disk. Moreover, 
\citet{Chern06} have found 
remarkable hardening down to $\Gamma \simeq 1.2$ 
at the entrance of the disk prior to periastron, 
followed by a recovery to $\Gamma \simeq 1.5$ during the disk crossing. 

In the hard X-ray domain, 
a dedicated  three-week \emph{CGRO} observation in 1994 
has resulted in a significant detection in the 30--200 keV range 
at a few mCrab level with the OSSE instrument, with the spectrum 
being characterized by a power law of $\Gamma = 1.8\pm 0.6$ \citep{Grove95}. 
The OSSE spectrum was consistent with the 
time-averaged \emph{ASCA} spectrum extrapolated to the OSSE band. 
With the \emph{INTEGRAL} satellite, the hard X-ray emission 
in the 20--200 keV band was detected with $\sim 4\sigma$ significance based 
on the observation performed in 21--25 March 2004 \citep{Shaw04}.

The properties of  the X-ray and soft $\gamma$-ray 
emission observed with \emph{ASCA} and \emph{CGRO}  
have been shown to be consistent with the idea that the emission is 
produced by shock-accelerated \pair\ pairs as a result of 
the interactions between the pulsar wind and the Be star outflows 
\citep{TAK94,Kaspi95,TA97}. 
The \pair\ pairs are considered to be accelerated at the inner shock 
front of the pulsar wind  and 
adiabatically expanding in the relativistic flow of the pulsar cavity.
As a scale-down version of pulsar wind nebulae (PWNe), 
the \PSR\ system can be referred to as a {\it compactified pulsar wind nebula}, 
and serves as a unique diagnostic of shock acceleration subject to 
very fast cooling. 
Indeed, synchrotron radiation by the accelerated \pair\ pairs offers an 
excellent explanation for the observed X-ray emission, indicating 
 the existence of a remarkably efficient acceleration mechanism 
at an energy of TeV on a timescale less than $\sim 100$ sec 
\citep{TA97}. 
The anisotropic inverse-Compton scattering on the intense stellar light 
of the accelerated \pair\ pairs was predicted to be detectable 
by air Cherenkov telescopes \citep{Kirk99}.

During the periastron passage in 2004, 
very high energy gamma-rays have been indeed detected by 
the HESS telescopes \citep{HESS05}.
The level of the TeV gamma-ray flux and the spectral shape 
were in a reasonable agreement with the theoretical expectation 
based on the compactified PWN model. 
However, the light curve around periastron disagreed with 
the early expectation, requiring  a revision of the model\footnote{Even departing 
from the  framework of the compactified PWN, 
a hadronic origin of TeV gamma-rays 
 \citep{Kawachi04} may replace the synchrotron-IC model 
 \citep[see also][]{Chern06}.}.
\citet{Mitya07} have introduced the change of the maximum energy of pairs 
as a viable explanation for the behavior of the TeV light curve. 
Also, they argued that 
the increase of the adiabatic loss rate near periastron is another possible 
way to naturally account for the reduction of the TeV flux. 
Since the change of the maximum energy and/or the adiabatic loss rate 
can be imprinted in the 
synchrotron X-ray spectrum, the detailed X-ray observations are thus 
quite important to understand the behavior of the TeV radiation 
as emphasized by \citet{Mitya07}. 

In this paper, we present the results from our monitoring observations 
made with the \emph{Suzaku} satellite in the X-ray and hard X-ray bands, 
with an effective energy range of 0.6--50 keV, 
performed during the most recent periastron passage in 2007.
Combined with the previous measurements in X-ray, 
the \emph{Suzaku} observations have clarified the spectral evolution 
of the \PSR\ system. 
Thanks to the wide band coverage, 
we have discovered the existence of a spectral break  in a certain epoch.
This paper is organized as follows. 
The observations and basic data reduction are described in \S\ref{sec:obs}.
The results of X-ray analyses are presented in \S\ref{sec:ana}.
In \S\ref{sec:interpretation}, we outline our model of 
a compactified pulsar wind nebula, and 
interpret the X-ray spectral break as 
a manifestation of the Lorentz factor of the pulsar wind.  
Throughout this paper, we adopt the distance to the \PSR\ system from 
the Earth to be $D = 1.5\ \rm  kpc$ \citep{Johnston94}.
For the calculation of the pulsar orbit, we assume 
a mass of SS~2883 to be $M_{\rm c} = 10M_\sun$ and 
the pulsar mass to be $M_{\rm p} = 1.4M_\sun$, which implies 
an inclination angle of $i \simeq 36\degr$ \citep{Johnston94}. 
The adopted orbital elements of the pulsar 
are an orbital period of 1236.72 day, 
eccentricity of $e = 0.8699$, longitude of periastron of $\omega = 138.7\degr$,
and the epoch of periastron of MJD 48124.34 \citep{Manchester95}.
All errors are quoted at a 90\% confident level unless otherwise stated.

\section{Observations and Data Reduction}\label{sec:obs}

We observed the PSR B1259$-$63 system eight times 
with the \emph{Suzaku} satellite each with $\sim 20$ ks exposure 
in 2007 July, August, and September. 
The total exposure time amounts to $\simeq 170$ ks 
after standard data screening. 
We note that this is the first and possibly last chance for 
\emph{Suzaku} to observe the first disk transit 
since target visibility does not permit us to observe it next time. 
Table~\ref{tbl:suzakulog} gives the log of the \emph{Suzaku} observations;
the eight observing epochs are referred to as Sz1--Sz8 hereafter.
During the first four epochs (Sz1--Sz4) 
simultaneous observations with HESS were coordinated. 
The results of the HESS observations will be presented separately. 

The \emph{Suzaku} observations were performed with the 
X-ray Imaging Spectrometer \citep[XIS:][]{XIS07} in 0.3--12 keV 
and 
the Hard X-ray Detector \citep[HXD:][]{Takahashi07} 
in 13--600 keV. 
The XIS, located at the focal plane of the X-ray telescopes 
\citep[XRT:][]{XRT07},  consists of one back-illuminated 
CCD camera (XIS-1) and three front-illuminated CCDs (XIS-0, -2 and -3).
One of the front-illuminated CCDs, 
XIS-2, was not available at the time of our observations, 
since it suffered from a fatal damage on 2006 November 9,
and unusable since then. 
The HXD consists of the silicon PIN photo diodes (hereafter PIN) 
capable of observations in the 13--70 keV band and the 
GSO crystal scintillators (hereafter GSO) which cover the 40--600 keV band. 

The XIS instruments were operated in a normal full-frame clocking mode, 
with a frame time of 8 s. 
In this work, we present results from front-illuminated CCDs
(XIS-0 and -3), which have a larger 
effective area at high energies, and therefore suited for our purpose. 
 XIS-0 and -3 have almost identical properties, 
so we co-added the data to improve photon statistics. 
\PSR\ was observed slightly off-axis ($3.5\arcmin$ from the optical axis), 
placed on the HXD nominal pointing position, to maximize the 
effective area of the HXD detectors. 
The HXD instruments were operated in the normal mode. 
The relative normalization of  the HXD-PIN spectrum to the XIS 
was fixed at 1.15,  as was determined by the calibration observations of the 
Crab Nebula \citep{Kokubun07}. 

\begin{figure}[h]
\epsscale{0.9}
\plotone{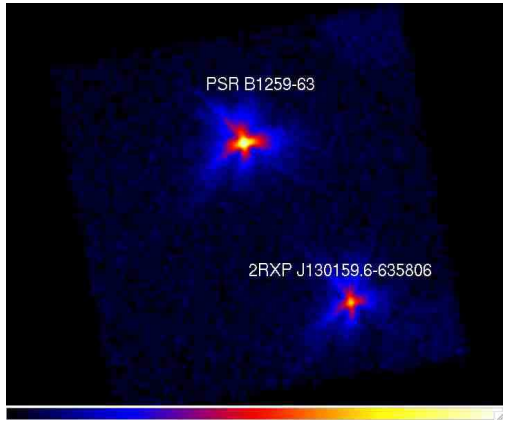}
\caption{\small  \emph{Suzaku} XIS0 1--10 keV image (epoch Sz1), 
shown with square root scaling. 
The FOV of the XIS is $17.8\arcmin \times 17.8\arcmin$. 
An accretion-powered pulsar \RXP\ is located $9.6\arcmin$ away from \PSR.
\label{fig:image} }
\end{figure}

The field-of-view (FOV) of the XIS contains another X-ray source, 
2RXP J130159.6$-$635806 (shortly \sRXP), 
as shown in Figure \ref{fig:image}. 
This is 
an accretion-powered pulsar with a spin period of $\sim 700$~s 
\citep{Chern05}, which is located $10\arcmin$ away from 
PSR B1259$-$63. 
We have  monitored the 700-sec accretion-powered pulsar with the XIS 
by tuning a telescope roll angle, 
to estimate its contribution to the hard X-ray flux 
measured by the non-imaging HXD-PIN detector. 

We analyzed data delivered to us via pipeline processing 
version 2.0.6.13 (Sz1--Sz5) or version 2.1.6.15 (Sz6--Sz8).
Basic  analysis was done using the HEASOFT software package (version 6.5), 
with calibration files distributed on 9 July 2008. 
For the XIS and HXD-PIN, 
we made use of cleaned event files, in which standard screening was applied. 
On the other hand, we reprocessed the HXD-GSO data by ourselves 
to apply appropriate gain files. 
The standard screening procedures include event grade selections, and 
removal of time periods  such as 
spacecraft passage of  the South Atlantic Anomaly (SAA), 
intervals of small geomagnetic cutoff rigidity (COR), 
and those of a low elevation angle. 
Specifically, for the XIS, 
the elevation angle 
larger than $5\degr$ above the Earth 
and  larger than $20\degr$ from the sunlit Earth limb are selected. 
The data screening criteria for the HXD include 
the COR larger than $6\ {\rm GV}$, and 
the elevation angle above the Earth limb greater than 5$\degr$. 

\begin{deluxetable}{ccccll}
\tabletypesize{\small}
\tablecaption{Log of \emph{Suzaku} Observations\label{tbl:suzakulog}}
\tablewidth{0pt}
\tablehead{
\colhead{ID} & \colhead{Date} & \colhead{MJD} & 
\colhead{$\tau$} & \multicolumn{2}{c}{Exposure (ks)} \\
\colhead{} & \colhead{} & \colhead{} & 
\colhead{(days)} &  \colhead{XIS}  & \colhead{PIN} 
}
\startdata
Sz1 & 2007 07 07 &  54288.6 & $-19.3$  & 21.9  &  25.3\\
Sz2 & 2007 07 09 &  54290.7 & $-17.2$  & 19.5 &  25.4\\
Sz3 & 2007 07 11 &  54292.6 & $-15.3$  & 22.7  &  22.9\\
Sz4 & 2007 07 13 &  54294.7 & $-13.2$  & 22.9  &  19.7\\
Sz5 & 2007 07 23 &  54304.3 & $-3.6$   & 19.7  &  16.7\\
Sz6 & 2007 08 03 &  54315.3 & $+7.4$  & 24.0  &  20.1\\
Sz7 & 2007 08 18 &  54330.1 & $+22.2$ & 20.5  & 18.1\\
Sz8 & 2007 09 05 &  54348.2 & $+40.3$  & 18.3  &  20.8 
\enddata
\tablecomments{
Modified Julian Date (MJD) and days from periastron ($\tau$) refer to 
the start time of the observation. 
The exposure time is the 
net integration time after standard data screening.}
\end{deluxetable}

\section{Analysis and Results}\label{sec:ana}

\subsection{Power-law fit to XIS spectra}

\begin{deluxetable}{ccccl}
\tabletypesize{\small}
\tablecaption{Results of \PSR\ Suzaku XIS Spectral Fitting \label{tbl:XISparameter}}
\tablewidth{0pt}
\tablehead{
\colhead{ID} & \colhead{$N_{\rm H}$} & \colhead{$\Gamma$} & \colhead{$F_{1-10}$} 
& \colhead{$\chi^2_\nu ({\rm d.o.f.})$}\\
\colhead{} & \colhead{$10^{22}\ \rm cm^{-2}$} & \colhead{} & 
\colhead{$10^{-12}\ \rm erg\ cm^{-2}\ s^{-1}$} & \colhead{}
}
\startdata
Sz1 & $0.50\pm 0.02$ & $1.64\pm 0.02$ & $25.8\pm 0.3$ & 1.06 (354)\\
Sz2 & $0.51\pm 0.03$ & $1.58\pm 0.03$ & $26.7\pm 0.3$ & 1.04 (309)\\
Sz3 & $0.51\pm 0.03$ & $1.35\pm 0.03$ & $21.6\pm 0.3$ & 1.19 (272)\\
Sz4 & $0.52\pm 0.03$ & $1.44\pm 0.03$ & $23.1\pm 0.3$ & 1.30 (298)\\
Sz5 & $0.50\pm 0.03$ & $1.83\pm 0.03$ & $19.0\pm 0.2$ & 0.85 (249)\\
Sz6 & $0.52\pm 0.03$ & $1.73\pm 0.03$ & $12.9\pm 0.2$ & 0.92 (205)\\
Sz7 & $0.50\pm 0.02$ & $1.69\pm 0.02$ & $34.4\pm 0.4$ & 1.01 (402)\\
Sz8 & $0.46\pm 0.03$ & $1.57\pm 0.03$ & $22.7\pm 0.3$ & 1.12 (253)
\enddata
\tablecomments{Fitting \emph{Suzaku} XIS spectrum of \PSR\ 
by a power law with photoelectric absorption 
in 0.6--10 keV. 
Absorbing column density $N_{\rm H}$, 
 photon index $\Gamma$, and the 1--10 keV flux $F_{\rm 1-10}$ 
 (not corrected for absorption) are  shown  with 90\% errors.}
\end{deluxetable}

The XIS spectra were extracted from circular regions of $3\arcmin$ radius
centered on the source. 
Background spectra were extracted from source free regions. 
Response matrices (``rmf") were generated by {\tt xisrmfgen}, and 
ancillary response functions (``arf") were simulated by running 
{\tt xissimarfgen}.
The source count rate is typically 1 counts s$^{-1}$ per XIS in 0.6--10 keV,
while the background 
count rate is only about $3\times 10^{-2}$ counts s$^{-1}$.
Source spectra were binned to a minimum of 100 counts per bin.
We performed spectral fitting in a range of 0.6--10 keV 
after excluding a narrow bandpass around the Si-K edge 
(1.8--1.9 keV) where some calibration errors remain.  

Given that  no previous observations have reported 
spectral changes within a day,
we first performed model fitting of epoch-by-epoch 
X-ray spectra of  the \PSR\ system accumulated for $\sim 20$ ks. 
 As expected from past X-ray observations \citep[e.g.,][]{Kaspi95,Chern06}, 
all the XIS spectra of the \PSR\ system show a featureless continuum without 
any emission lines.
The \emph{Suzaku} X-ray spectra 
are shown later in Figure~\ref{fig:xispin}.
The XIS spectrum (in terms of number flux per energy) was modeled by 
a power law, 
namely $F(\epsilon)=K \epsilon ^{-\Gamma}$ 
where 
$K$ is the normalization and $\Gamma$ is the photon index, 
attenuated by an absorption  factor of 
$\exp [- N_{\rm H} \sigma (\epsilon)]$. 
Here  $N_{\rm H}$ is 
the absorption column density and $\sigma (\epsilon )$ represents 
the photoelectric absorption cross section. 
The results of the power-law fit are 
summarized in Table~\ref{tbl:XISparameter}. 
The best-fit photon indices are found in a wide range of 
$\Gamma = 1.35\mbox{--}1.83$, while 
the absorption column density in a narrow range of 
 $N_{\rm H} = (0.46\mbox{--}0.52)\times 10^{22}\ \rm cm^{-2}$.
In most cases we obtained a statistically acceptable fit. 
However, in the case of Sz4,  the absorbed power-law fit was 
unacceptable with a reduced chi-square of $\chi_{\nu}^2 = 1.30$ for 298 
degree of freedom (d.o.f.). 
Also, the fitting  quality for Sz3 seems bad with 
$\chi_{\nu}^2 = 1.19$ for 272 d.o.f. It should be noted that 
a  quite hard power law was obtained in both epochs. 
Inspection of the fit residuals suggests the presence of spectral steepening.
This issue shall be investigated in detail using the HXD-PIN in \S\ref{sec:broadband}.

\begin{figure*}[htbp]
\epsscale{0.9}
\plotone{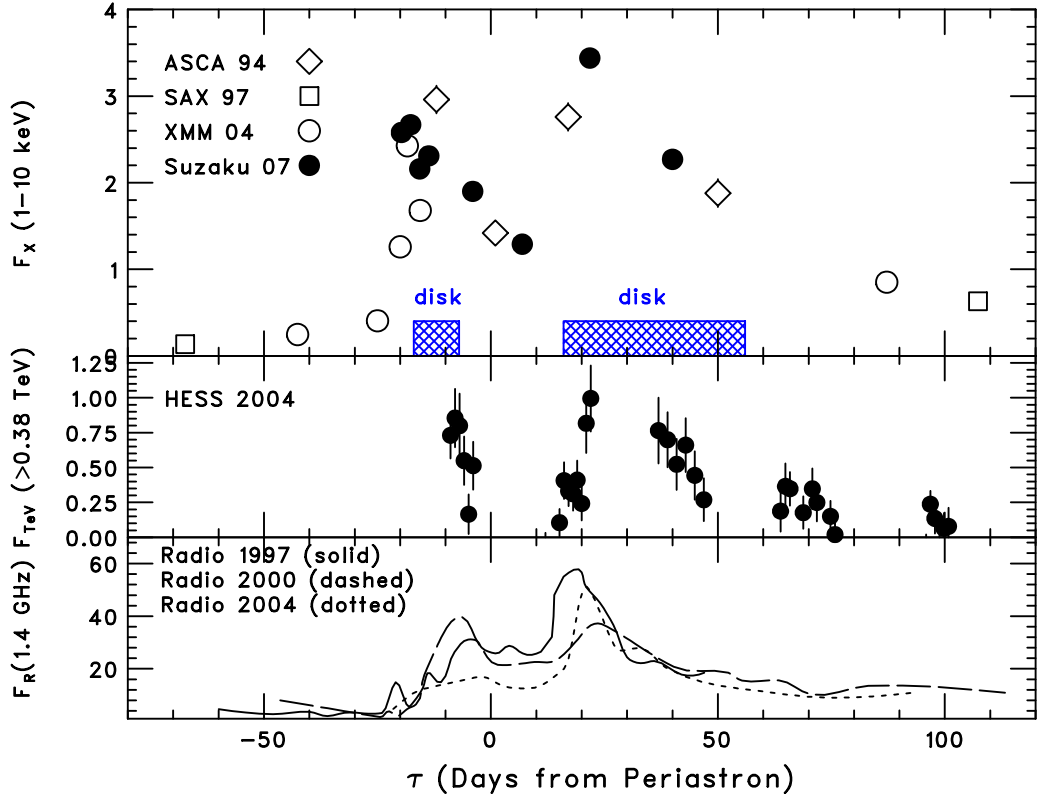}
\caption{\small  
X-ray, TeV, and radio fluxes as a function of 
time from periastron passage. ({\it Top}) 1--10 keV flux lightcurve 
in units of $10^{-11}\ \rm erg\ cm^{-2}\ s^{-1}$. 
Absorption is not corrected for. 
Errors are small compared with the mark size. 
The \emph{Suzaku} results 
obtained in this work are drawn as filled circles. 
The \emph{ASCA} GIS data points in 1994 \citep{Hirayama96},
the \emph{Beppo-SAX} ones in 1997, 
 and the \emph{XMM} ones in 2004 \citep{Chern06} are shown as 
open triangles, squares and circles, respectively. 
The periods over which the pulsar crossed the equatorial disk are 
shown (blue hatched regions) based on the geometry proposed by \citet{Chern06}.
({\it Middle}) Gamma-ray flux above 0.38 TeV  in  units of 
 $10^{-11}\ \rm cm^{-2}\ s^{-1}$ obtained with HESS \citep{HESS05}.
 ({\it Bottom}) 1.4 GHz radio lightcurves in 1997, 2000, 2004 \citep{Johnston05}.
\label{fig:suzaku_hess_radio} }
\end{figure*}

Figure~\ref{fig:suzaku_hess_radio} 
plots the  flux measured by the \emph{Suzaku}  XIS 
as a function of $\tau$ (days from periastron), 
together with a compilation of previous measurements. 
The new \emph{Suzaku} data points largely 
help us track the 
time evolution of the X-ray flux from the \PSR\ system, and 
make it clear that 
the stellar disk  plays essential roles in regulating the 
X-ray emission from the \PSR\ system.
In the X-ray lightcurve as a function of $\tau$ (Fig.~\ref{fig:suzaku_hess_radio}),
there appear two bumps, which can be identified with the disk passages 
of the pulsar.
The position of the post-periastron peak at epoch Sz7 ($\tau = +22$ days) 
coincides with the enhancement of the TeV flux measured with 
HESS in 2004, as well as the peak in the radio lightcurves. 
In accordance with the disk geometry introduced by \citet{Chern06}, 
we associate this feature with the pulsar's re-entrance 
to the Be star equatorial  disk. 
The relatively smooth multi-orbital 
lightcurve over a decade suggests that orbit-by-orbit differences 
in X-ray behavior would be small, though the current data sets are 
not enough to draw a definitive conclusion. 
We note that the unpulsed radio lightcurves show significant 
differences from one periastron to next (see Fig.~\ref{fig:suzaku_hess_radio}), 
which would support the idea that 
the radio electrons originate from the shocked equatorial disk \citep{Ball99}.

\begin{figure}[htbp]
\epsscale{1.1}
\plotone{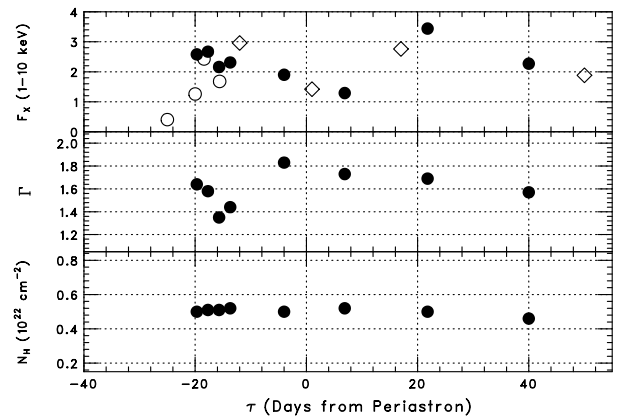}
\caption{\small  \emph{Suzaku} XIS best-fit parameters as a function of 
time from periastron passage. Statistical errors are smaller than the 
size of marks. 
({\it Top}) 
The same lightcurve as the top panel of Fig.~\ref{fig:suzaku_hess_radio}.
({\it Middle}) Photon index $\Gamma$. 
({\it Bottom}) Absorption column density $N_{\rm H}$. 
\label{fig:tau_param} }
\end{figure}

In Figure~\ref{fig:tau_param}, the best-fit values of  $\Gamma$ 
and $N_{\rm H}$ are shown as a function of $\tau$.
As mentioned above, the column density is found to be almost constant,
 $N_{\rm H} \simeq 0.5 \times 10^{22}\ \rm cm^{-2}$, 
 over the course of \emph{Suzaku} monitoring. 
This is roughly consistent with the values obtained  with 
\emph{XMM-Newton} 
during the previous disk passage in 2004, 
 $N_{\rm H} \simeq 0.45 \times 10^{22}\ \rm cm^{-2}$  \citep{Chern06}.
It should be noted that smaller column density
($N_{\rm H} \sim 0.3 \times 10^{22}\ \rm cm^{-2}$)   
 has been observed 
with \emph{XMM-Newton} when the pulsar was positioned outside the disk
\citep{Chern06}.
The excess column density can therefore be ascribed to the dense stellar disk. 
On the other hand, photon index exhibits remarkable changes; 
significant spectral flattening characterized by $\Gamma \simeq 1.35$ was observed 
during the first disk transit ($\tau \sim -15\ \rm d$).
Such spectral hardening was discovered in 2004 with \emph{XMM-Newton}, 
but it appeared  slightly earlier in phase ($\tau \sim -20\ \rm d$).
Later in section \S\ref{sec:broadband},  
we will demonstrate the presence of 
a spectral break during the hard-spectrum state, 
by utilizing hard X-ray measurements 
in the 15--60 keV band with the \emph{Suzaku} HXD-PIN.

\subsection{Lightcurves} 

\begin{figure}[htbp]
\epsscale{1.1}
\plotone{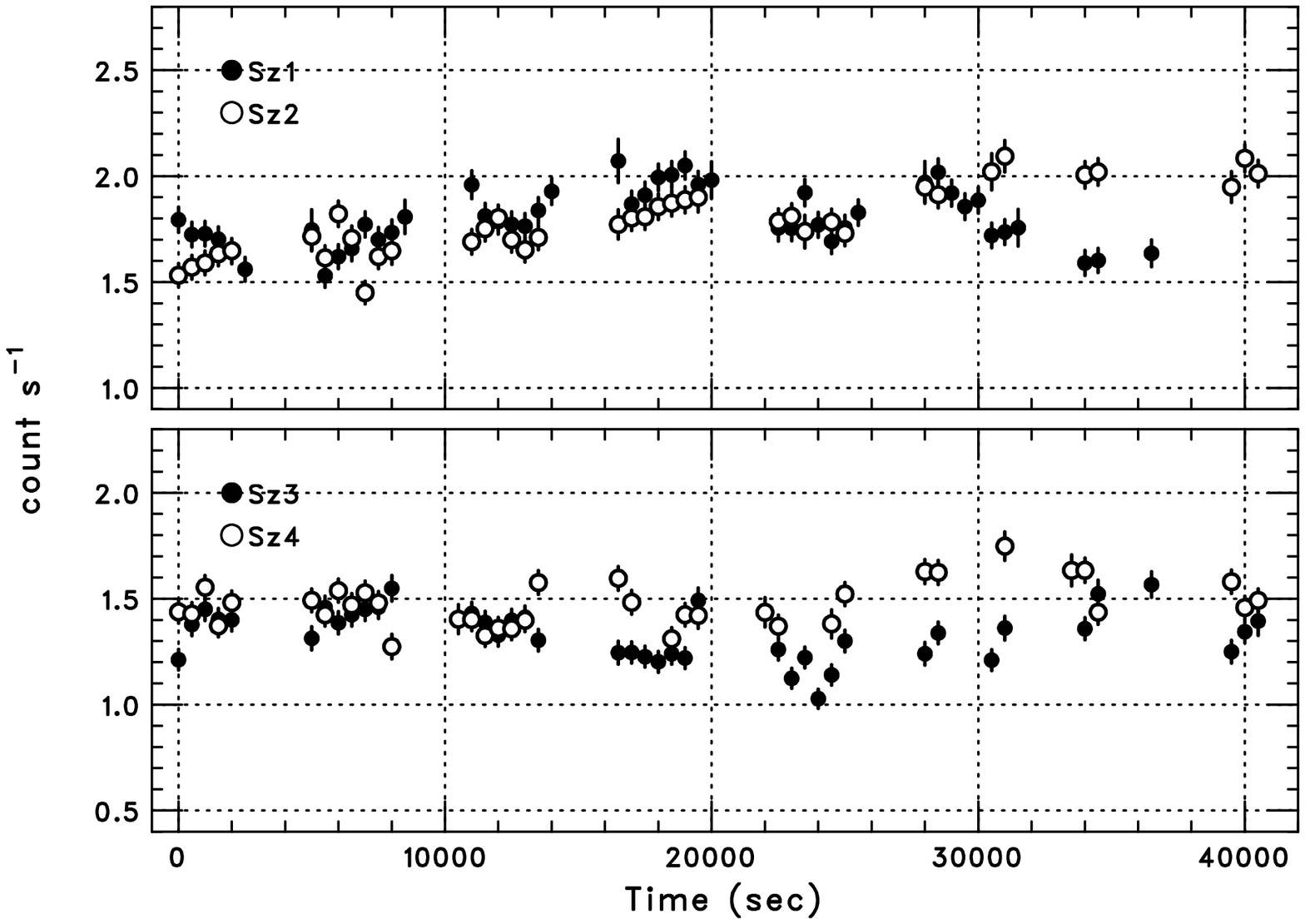}
\plotone{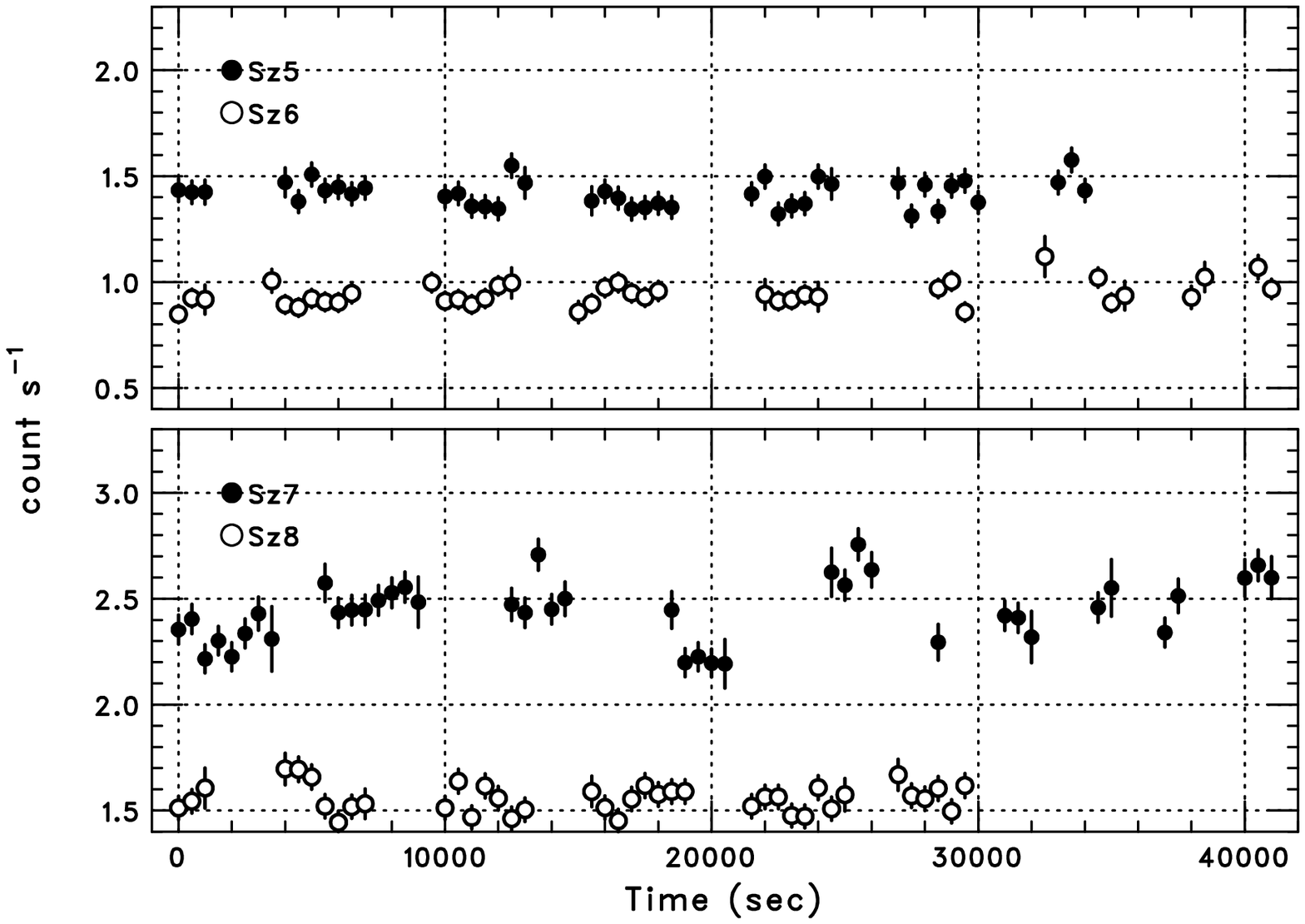}
\caption{\small  \emph{Suzaku} lightcurves of \PSR\  in the 1--10 keV band.
The count rate with a bin size of 500 s 
from the start time of the observation 
is shown without background 
subtraction, whose contribution is  only a few percent. 
\label{fig:rate} }
\end{figure}

The X-ray lightcurves 
for individual observations were examined in order to study 
the temporal characteristic of the X-ray emission on short time scales.
The 1--10 keV lightcurves with a bin size of 500 sec are shown in 
Figure~\ref{fig:rate}. 
We found moderate hour-scale flux variability ($\sim 30\%$) 
at  the time of disk crossings.
On the other hand, 
the lightcurves in epochs Sz5 and Sz6 can be fit well with a constant, 
suggesting that the flux variations would be characteristics of disk passage.
The size of the emitting regions may be smaller in the equatorial disk.
Alternatively, there is a large density fluctuation inside the disk. 

To search for possible spectral changes within each observation, 
we subdivided each epoch into several periods typically with an hour exposure, 
and fitted the short time-frame spectra with a power law modified by
photoelectric absorption. 
Though the power-law  normalization differs between sub-epochs, 
no statistically significant spectral changes  were found, 
which indicates the stability of  a spectral shape  on hour scales.

The density fluctuation inside the Be star disk would be imprinted in 
the chaotic behavior of the X-ray lightcurves on hour scale 
in one way or another. 
The hour-scale variability \emph{Suzaku} uncovered would offer a way to infer 
the pulsar's location relative to the Be disk plane.
Future intense coverages during the post-periastron disk transit of the pulsar 
with \emph{Suzaku} will be able to constrain the structure of the Be star disk.

\subsection{Preparation for Broadband Analysis}\label{sec:prep}

With the \emph{Suzaku} observations, we are now accessible to 
 a {\it day-by-day} broadband X-ray spectrum 
spanning two orders of magnitude in energy from 0.6 to 50 keV
 with sufficient sensitivities, by combining the XIS (0.6--10 keV) and 
 the HXD-PIN (15--50 keV) instruments. 

To extract the HXD-PIN spectra of \PSR, we have to subtract 
the following background spectra from the total cleaned spectra:
(1) instrumental background so-called non X-ray background (NXB), 
(2) cosmic X-ray background (CXB), that is the sum of unresolved active galactic nuclei 
in the FOV of the PIN, and (3) the emission from the nearby pulsar 
\sRXP\ that lies within the FOV of the PIN. 
The instrumental background spectra (perhaps due mostly to atmospheric neutrons), 
provided by the HXD instrument team, 
are constructed by  the empirical models, which rely on 
 a database of Earth occultation observations. 
The HXD team provides two kinds of 
background models, ``quick" and ``tuned".
We found the two models give almost identical results within the statistical uncertainties.
In this paper we make use of  the ``tuned" NXB model, which  offers better 
background reproducibility achieving the $1\sigma$ systematic error of $<$3\% 
\citep{Fuka08}. The systematic error of the NXB subtraction does not have 
noticeable impact on the results presented in this paper. 

\begin{figure}[htbp]
\epsscale{1.1}
\plotone{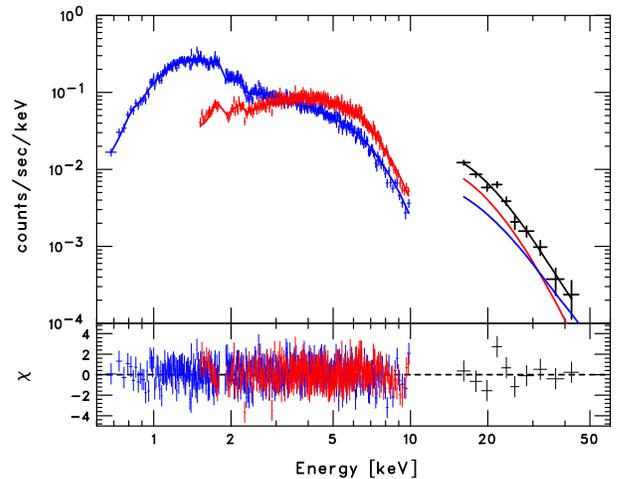}
\caption{\small  
\emph{Suzaku} XIS and PIN spectra obtained by 
combining epochs Sz5 and Sz6.
The XIS spectrum of \PSR\ is shown 
in blue, and that of  \sRXP\   in red. 
The PIN spectrum  (black crosses) 
after  background  subtraction 
is decomposed into the two components through joint fitting:
\PSR\ power-law model (blue line), and 
\sRXP\ cutoff power-law model (red line).
\label{fig:joint} }
\end{figure}

The CXB flux 
in the FOV of the PIN instrument was subtracted from each PIN spectrum 
in the same way as \citet{Takahashi08}.
The CXB flux amounts to  $\sim 5$\% of the NXB,
comparable to the systematic error of the NXB model itself.

We now describe the procedures used for the derivation of 
hard X-ray spectra of \sRXP\ in each pointing. 
In fact, the main contributions to the systematic error of the PIN spectral data 
of  the \PSR\ system itself 
would be the uncertainties in the hard X-ray flux contaminated by \sRXP. 
To describe the X-ray spectrum of \sRXP, 
we adopt the following empirical spectral form that has been traditionally 
applied to accretion-powered pulsars:
\begin{equation}
\label{eq:RXP}
F(\epsilon ) = \left\{
\begin{array}{ll}
K \epsilon^{-s} 
& \epsilon \leq \epsilon_{\rm c}  \\
K \epsilon^{-s}  \exp \left( -(\epsilon-\epsilon_{\rm c})/\epsilon_{\rm f}  \right)
& \epsilon > \epsilon_{\rm c},
\end{array}
\right.
\end{equation}
where $K$ is the normalization \citep{WSH83}. 
The index $s$ was frozen to be $s=1$, 
a typical value  for accretion-powered X-ray pulsars \citep[e.g.,][]{Nagase89}, 
 to reduce the number of free parameters. 
Noncyclotron pulsars can be characterized by 
$\epsilon_{\rm c} = 4\mbox{--}15\ \rm keV$ and 
$\epsilon_{\rm f} = 10\mbox{--}30\ \rm keV$ \citep{Maxima99}. 

We chose the epochs Sz5 and Sz6 
for the use of determination of  the spectral shape of \sRXP\ 
in the 1--50 keV band, because the relative contribution of \sRXP\ 
in a PIN spectrum would be large in the two epochs, 
where the \PSR\ emission is relatively weak and steep.  
To obtain the model parameters relevant for \sRXP, 
the XIS spectrum of \PSR\ in 0.6--10 keV, that of \sRXP\ in 1.5--10 keV, and 
the PIN spectrum in 15--50 keV 
after subtraction of the background (NXB+CXB) were jointly fit 
by a two component model consisting of a power law 
that describes \PSR, and equation (\ref{eq:RXP}) that describes  \sRXP.
Each spectral component is attenuated by photoelectric absorption. 
The spectral data for the two epochs were summed to reduce 
statistical errors, because the XIS spectra of \PSR\ (and also \sRXP ) 
are  similar in the two epochs. 
An energy-independent transmission factor (about 0.7) 
of the PIN collimator at the position of 
\sRXP\ was taken into account when performing a joint fit.
The assumption of a  power-law spectrum  for \PSR\ in the 0.6--50 keV band 
in epochs Sz5 and Sz6 
can be justified, to some extent, 
by the OSSE results \citep{Grove95}, which came from the three-week 
observation covering 
a certain orbital phase that encompasses the two \emph{Suzaku} epochs. 

The joint fit was statistically acceptable with  a reduced chi-square of 
$\chi_{\nu}^2 = 1.02$ for 768 d.o.f. 
In Figure~\ref{fig:joint} the XIS spectra of both \PSR\ and \sRXP, and 
the PIN spectrum (the sum of \PSR\ and \sRXP\ contributions) are 
shown together with the best-fit model. 
We obtained the following parameters for the \sRXP\ model:
$N_{\rm H} = (2.62  \pm 0.08) \times 10^{22}\ \rm cm^{-2}$, 
$\epsilon_{\rm c} = 4.4^{+0.3}_{-0.4}\ \rm keV$, and 
$\epsilon_{\rm f} = 14.1 \pm 1.2\ \rm keV$.
The values seem reasonable  for noncyclotron pulsars.
The best-fit parameters for the \PSR\ model are 
$N_{\rm H} = (0.50  \pm 0.02) \times 10^{22}\ \rm cm^{-2}$, and 
$\Gamma = 1.77 \pm 0.02$, which are consistent with 
the results obtained using the XIS alone.

The \emph{Suzaku} XIS spectrum of \sRXP\  
at each epoch was then fit  by using  the cutoff power-law model  with 
the parameters obtained above 
(namely, $\epsilon_{\rm c} = 4.4\ \rm keV$ and $\epsilon_{\rm f} = 14\ \rm keV$), 
 to estimate the flux contamination of \sRXP\ in each PIN spectrum.
Spectral fitting was performed in an energy range of 2--10 keV after excluding 
an iron-K band of 6.2--6.8 keV in some occasions where we found possible iron lines. 
The template model with the fixed spectral shape 
gives an acceptable fit  in all cases, 
indicating the stability of  the spectral shape of \sRXP\  from one epoch to another. 
The flux level of \sRXP\ was also found to be stable, 
with the unabsorbed 1--10 keV flux  in a range of 
$F = (2.6\mbox{--}3.8) \times 10^{-11}\ \rm erg\ cm^{-2}\ s^{-1}$.

We also fit \emph{Suzaku} XIS spectrum of \sRXP\  at each epoch 
 by using an absorbed power-law model, instead of the cutoff power-law model of 
equation (\ref{eq:RXP}), to quantify spectral stability. 
The single power-law model gives an acceptable fit in the 1--10 keV band in all cases, 
as summarized in Table \ref{tbl:RXPparam}.
The observed spectral shape in this band looks  similar to each other.
The best-fit photon indices in epochs Sz1--8 are found to 
be  $\Gamma \simeq 1.2 \pm 0.1$, 
with the absorption column density of 
 $N_{\rm H} = (2.7\mbox{--}3.0)\times 10^{22}\ \rm cm^{-2}$.
 The spectral similarity from one epoch to another supports our assumption 
 that the parameter sets of $\epsilon_{\rm c}$ and $\epsilon_{\rm f}$ 
 obtained in Sz5+6 are applicable to other epochs as well. 
\begin{deluxetable}{ccccl}
\tabletypesize{\small}
\tablecaption{Results of \sRXP\ Suzaku XIS Spectral Fitting \label{tbl:RXPparam}}
\tablewidth{0pt}
\tablehead{
\colhead{ID} & \colhead{$N_{\rm H}$} & \colhead{$\Gamma$} & \colhead{$F_{1-10}$} 
& \colhead{$\chi^2_\nu ({\rm d.o.f.})$}\\
\colhead{} & \colhead{$10^{22}\ \rm cm^{-2}$} & \colhead{} & 
\colhead{$10^{-12} \rm erg\ cm^{-2}\ s^{-1}$} & \colhead{}
}
\startdata
Sz1 & $2.78^{+0.27}_{-0.25}$ & $1.20^{+0.10}_{-0.09}$ & $32.0^{+1.1}_{-1.0}$ & 1.09 (192)\\
Sz2 & $2.77^{+0.25}_{-0.24}$ & $1.18\pm {0.09}$ & $32.4^{+1.1}_{-1.0}$ & 0.91 (250)\\
Sz3 & $2.69^{+0.22}_{-0.20}$ & $1.19\pm {0.08}$ & $36.4\pm {1.0}$ & 0.95 (315)\\
Sz4 & $2.91^{+0.23}_{-0.21}$ & $1.20\pm {0.08}$ & $38.4^{+1.1}_{-1.0}$ & 1.10 (334)\\
Sz5 & $2.77^{+0.23}_{-0.22}$ & $1.18^{+0.09}_{-0.08}$ & $32.7^{+1.0}_{-0.9}$ & 1.02 (267)\\
Sz6 & $2.88^{+0.26}_{-0.24}$ & $1.28\pm {0.09}$ & $26.4^{+0.9}_{-0.8}$ & 1.03 (267)\\
Sz7 & $2.71^{+0.25}_{-0.23}$ & $1.17^{+0.09}_{-0.08}$ & $31.4^{+1.0}_{-0.9}$ & 1.05 (270)\\
Sz8 & $3.03^{+0.30}_{-0.27}$ & $1.26\pm {0.10}$ & $30.7^{+1.2}_{-1.1}$ & 1.00 (225)
\enddata
\tablecomments{Fitting \emph{Suzaku} XIS spectrum of \sRXP\ 
by a power law with photoelectric absorption 
in 1--10 keV. 
Absorbing column density $N_{\rm H}$, 
 photon index $\Gamma$, and unabsorbed 1--10 keV flux $F_{\rm 1-10}$ are 
shown  with 90\% errors.}
\end{deluxetable}

\subsection{Broadband Spectra}\label{sec:broadband}

\begin{figure}[htbp]
\epsscale{1.0}
\plotone{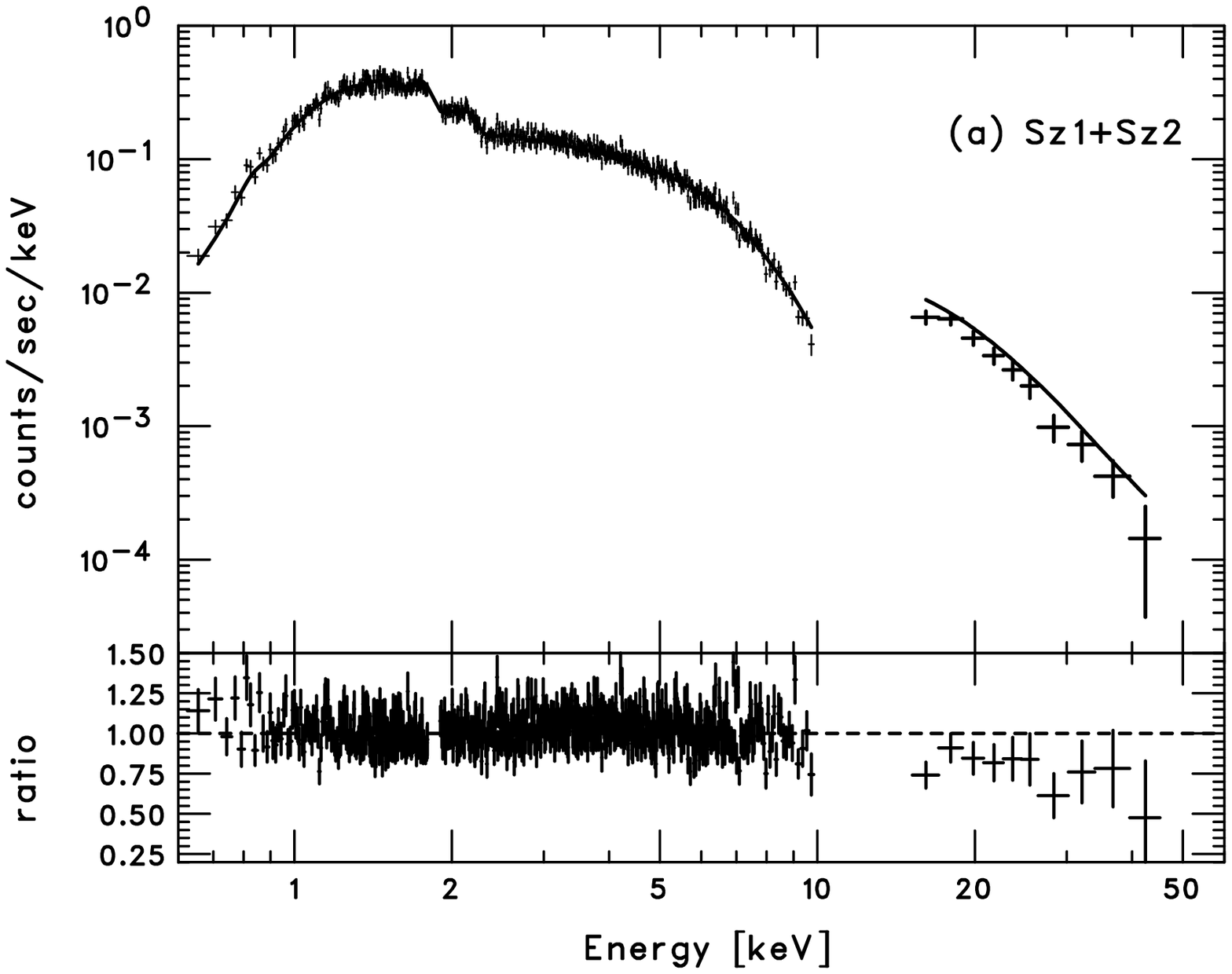}
\plotone{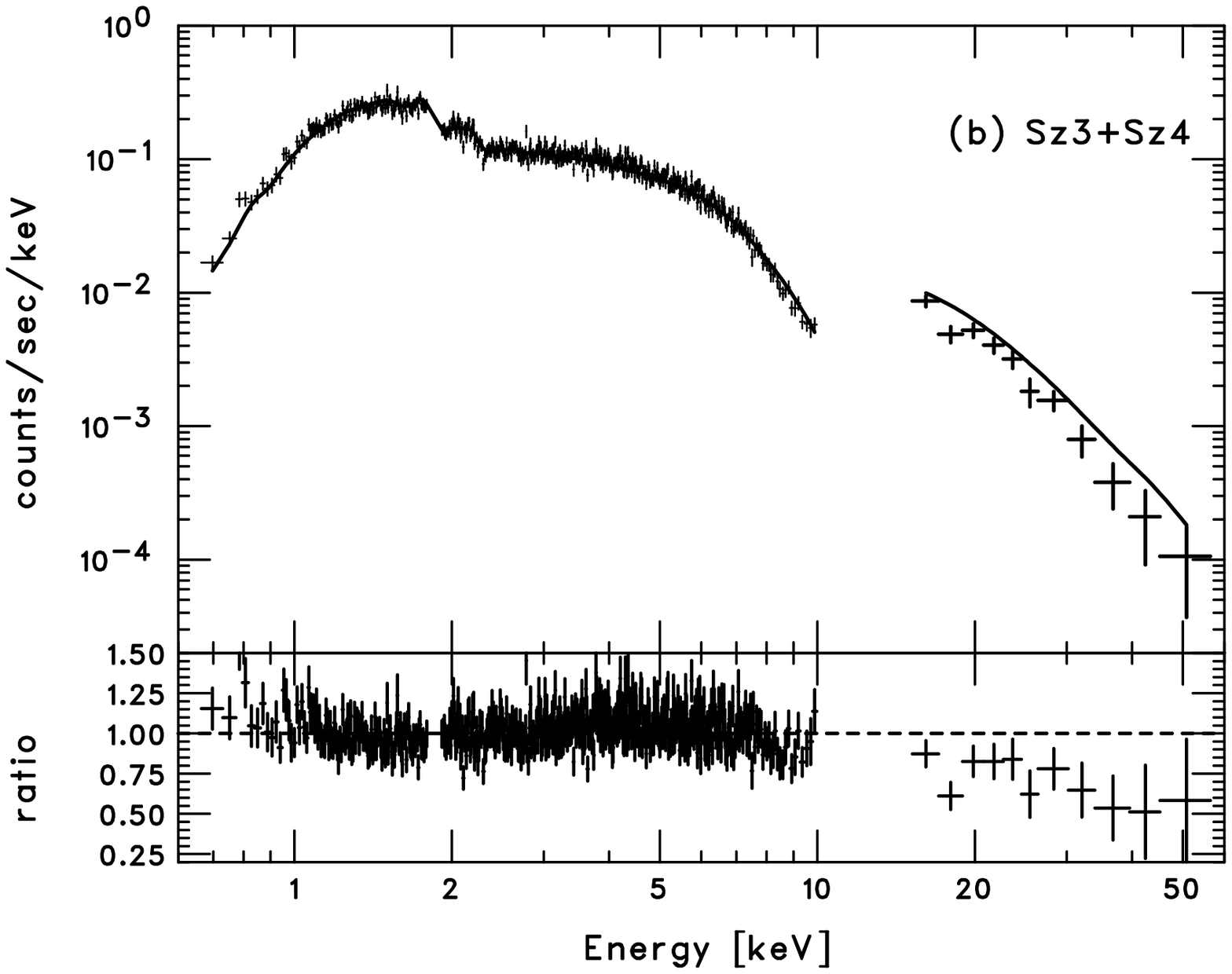}
\plotone{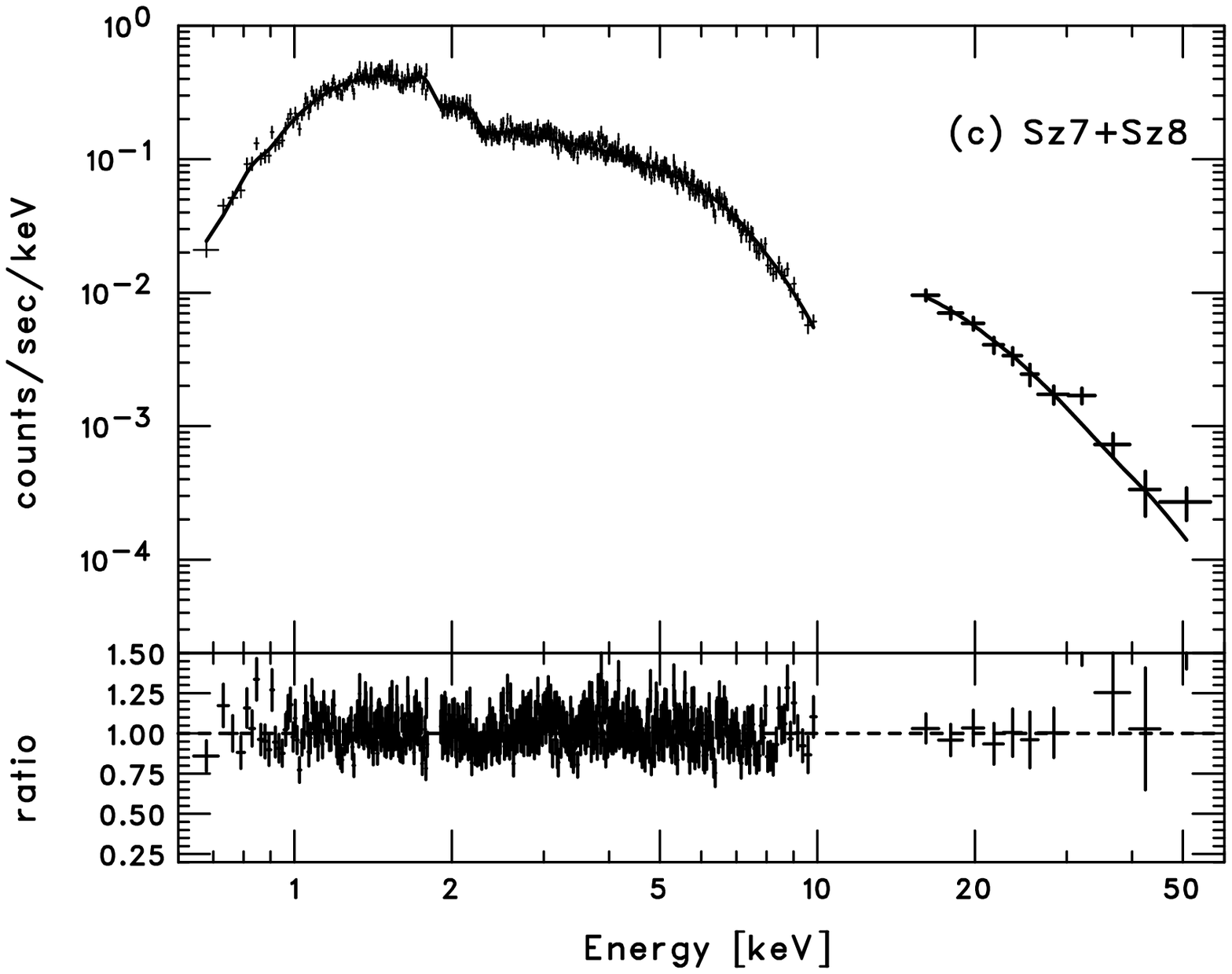}
\caption{\small 
{\it Suzaku} XIS+PIN spectrum of \PSR\ in the 0.6--50 keV band 
 and the ratio of the data and the best-fit absorbed power-law model:
 (a) Sz1+2, (b) Sz3+4, and (c) Sz7+8.
The contaminating flux from \sRXP\ in the PIN spectrum 
was subtracted in the way described in \S\ref{sec:prep}.
Superposed on the spectral data are the absorbed 
single power-law models, which do not provide a good fit 
to the Sz1+2 and Sz3+4 spectra.
\label{fig:xispin} }
\end{figure}

Figures \ref{fig:xispin} shows the \emph{Suzaku} 
XIS+PIN spectra of  the \PSR\ system.
Given the spectral similarity, 
two successive epochs were merged into a single observation, 
labeled as Sz1+2, Sz3+4, and Sz7+8. 
For the PIN spectrum, the NXB and CXB as well as 
the  contribution from \sRXP\ were subtracted (see \S\ref{sec:prep}).
We first performed spectral fitting using 
an absorbed single power law. 
The results are shown in Figure~\ref{fig:xispin} and the 
best-fit parameters are summarized in Table~\ref{tbl:XISPIN}.

The Sz3+4 spectrum could not be well described by a single power law;
there found wavy residuals in the XIS band, and, moreover, there exist 
flux deficits compared to the power-law model 
in the PIN spectra at high energies, indicating the presence of 
spectral steepening. 
It is interesting to note that similar wavy residuals to the best-fit power-law 
model are found in the XIS spectrum of the supernova remnant 
RX~J1713.7$-$3946, where the X-ray spectrum is indeed steepened 
\citep{Takahashi08}.
Imperfect subtraction of the \sRXP\ contamination is difficult to 
account for the flux deficit, because \sRXP\ has a much steeper spectrum 
in the PIN band and hardly affects the PIN spectrum above 30 keV. 
In fact, even if we artificially lower $\epsilon_{\rm f}$ to 10 keV  to 
increase the \sRXP\ contamination, the presence of spectral steepening 
is still required. 
We conclude that the Sz3+4 spectrum exhibits steepening 
toward high energies. 
Also, 
the Sz1+2 spectrum would 
require less significant steepening, the presence of which is not 
conclusive. 
The broadband Sz7+8 spectrum was found to be consistent with a 
simple power law with $\Gamma = 1.63 \pm 0.02$.

\begin{figure}[htbp]
\epsscale{1.1}
\plotone{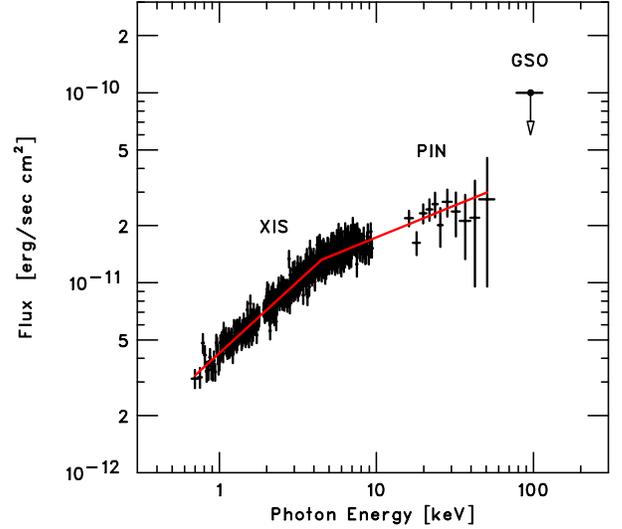}
\caption{\small  Broadband unfolded X-ray spectra of \PSR\ obtained 
by using the \emph{Suzaku} data in epochs Sz3 and Sz4.
The low-energy absorption is corrected for by the best-fit model. 
The best-fit broken power-law function is overlaid. 
\label{fig:nufnu} }
\end{figure}

We have proceeded to fit the Sz3+4 spectrum with a broken power-law model 
attenuated by photoelectric absorption, 
which yielded  an acceptable  fit (see Table~\ref{tbl:XISPIN}). 
The best-fit photon indices are 
 $\Gamma_1 = 1.25^{+0.02}_{-0.04}$ and 
$\Gamma_2 = 1.66^{+0.05}_{-0.04}$, with a break energy of 
 $\varepsilon_{\rm br} = 4.5 ^{+0.5}_{-0.2}$ keV. 
We also fit the Sz3+4 spectrum with a broken power-law model 
restricting the energy band to 0.6--10 keV. 
In such an XIS-only case, we obtained the following values: 
 $\Gamma_1 = 1.25\pm 0.04$,
$\Gamma_2 = 1.64\pm 0.04$, and $\varepsilon_{\rm br} = 4.4 \pm 0.2$ keV. 
Though the systematic error in the PIN spectrum itself would be large, 
its influence on the spectral parameters is therefore found to be small. 
Figure~\ref{fig:nufnu} presents the unfolded X-ray spectrum 
with the best-fit broken power-law model. 
 For reference, the broken power-law model was applied also to 
 the Sz1+2 and Sz7+8 spectra. As expected from the single power-law fitting, 
two photon indices for  the Sz7+8 spectrum are similar.

\begin{deluxetable*}{lcccccl}
\tabletypesize{\small}
\tablecaption{Results of \PSR\  XIS+PIN Spectral Fitting \label{tbl:XISPIN}}
\tablewidth{0pt}
\tablehead{
\colhead{ID} & \colhead{Model} & \colhead{$N_{\rm H}$} & \colhead{$\Gamma_1$} 
& \colhead{$E_{\rm br}$}
& \colhead{$\Gamma_2$} 
& \colhead{$\chi^2_\nu ({\rm d.o.f.})$}\\
\colhead{} & \colhead{} & \colhead{$10^{22}\ \rm cm^{-2}$} & \colhead{}  
& \colhead{keV} & \colhead{} & \colhead{}
}
\startdata
Sz1+2 & power law&   
$0.52\pm 0.02$ & $1.64\pm 0.02$ & \nodata & \nodata & 1.16 (604)\\
Sz1+2 & broken PL & 
$0.44^{+0.02}_{-0.03}$ & $1.48^{+0.05}_{-0.06}$
 & $3.7 \pm 0.5$ & $1.76\pm 0.04$  & 1.05 (602)\\
Sz3+4 & power law&   
$0.53\pm 0.02$ & $1.43\pm 0.02$ & \nodata & \nodata & 1.27 (500)\\
Sz3+4 & broken PL & 
$0.43 \pm 0.03$ & $1.25^{+0.02}_{-0.04}$
 & $4.5 ^{+0.5}_{-0.2}$ & $1.66^{+0.05}_{-0.04}$  & 0.99 (498)\\
Sz7+8 & power law&   
$0.48\pm 0.02$ & $1.63\pm 0.02$ & \nodata & \nodata & 1.12 (490)\\
Sz7+8 & broken PL &
$0.47 \pm 0.02$ & $1.59^{+0.04}_{-0.03}$
 & 4 (fixed) & $1.66^{+0.04}_{-0.03}$  & 1.11 (488)
\enddata
\tablecomments{Fitting \emph{Suzaku} XIS+PIN spectrum 
by a power-law or a broken power-law (PL) model 
with photoelectric absorption in 0.6--50 keV. 
Absorbing column density $N_{\rm H}$, 
 photon index $\Gamma_1$ (and $\Gamma_2$ in the case of broken PL), and 
the break energy $E_{\rm br}$ of a broken power-law,  are 
shown  with 90\% errors.}
\end{deluxetable*}

\subsection{Flux Upper Limit at 100 keV}

The HXD-GSO scintillation detectors cover the energy range of 50--600 keV.
Since the contribution of \sRXP\ is expected to be negligible above 50 keV, 
the measurement by the GSO is important for testing the presence of spectral 
steepening inferred by the XIS-PIN fitting. 
There found no significant detection 
in any epoch with the HXD-GSO.
We then set the upper limits on the source count rate 
as 2\% of the background count rate, 
which corresponds to a $\simeq 3\sigma$ limit. 
It should be noted that the uncertainty in GSO measurements
is dominated by the systematic error associated with background modeling. 
The upper limit of 
$1\times 10^{-10}\ \rm erg\ cm^{-2}\ s^{-1}$ was placed at 100 keV 
(in an energy range of 78--114 keV).
The limit depends only weakly on the assumed photon index ($\Gamma = 1.5$)
for a reasonable range of $\Gamma = 1\mbox{--}3$.
Since the averaged background level was almost unchanged during the  
 monitoring campaign, differences in
the upper limits from one epoch to another  are negligible 
(only $\sim 10\%$). 
Though the fact that the uncertainty comes from the systematic error 
prevents us from making a definitive statement about the significance, 
 the  upper limit set by the GSO 
 depicted in Fig.~\ref{fig:nufnu} 
strengthens the presence of spectral steepening.

\section{Interpretation of the Spectral Break}\label{sec:interpretation}

The most important finding with \emph{Suzaku} would be 
the presence of a remarkable spectral break 
by $\Delta \Gamma \simeq 0.4$ 
around $\varepsilon_{\rm br} \sim 5$ keV in the Sz3+Sz4 spectrum.
The low-energy part below $\varepsilon_{\rm br}$ can be described 
by a rather flat index of 
 $\Gamma_1 = 1.25\pm 0.04$. 
Here we consider the physical implication of this spectral 
structure within the framework of the shocked 
relativistic pulsar wind model, in which high-energy radiation 
comes from shock-accelerated electrons and positrons 
produced at wind termination shock \citep[e.g.,][]{TA97}.
Indeed, the synchrotron and IC radiation by the shock-accelerated 
\pair\ pairs 
from the \PSR\ system as results of the interaction between 
the relativistic pulsar wind and 
the Be star outflows have been studied extensively in previous 
work \citep{TA97,Kirk99,Murata03,Mitya07}, on which our considerations are based. 
However, if the particles in the relativistic wind can be efficiently mixed 
with the surrounding dense medium,  other radiation channels 
such as hadronic interactions may be  important 
\citep{Kawachi04,Chern06}; we do not consider such possibilities 
in this paper. 
Also, comprehensive modeling of the overall 
phenomenology is deferred to a future publication. 

\subsection{Model of Compactified Pulsar Wind Nebula}

Here we describe our shock-powered emission model, 
which is broadly similar to what 
has been devised in previous theoretical work mentioned above.
The \PSR\ system can be regarded as a scale-down version of PWNe. 
The relativistic wind of the young pulsar 
is presumed to be confined by the stellar outflows of the form of wind and disk, 
leading to the formation of termination shock inside the pulsar wind. 
The location of the termination shock should be close to the 
pulsar if the ram pressure of the stellar outflows 
as seen by the pulsar is large enough. The distance of the shock 
from the pulsar, $r_{\rm s}$, is assumed to be much smaller than 
the star-pulsar separation $d$: $r_{\rm s} \ll d$ \citep[e.g.,][]{Mitya07}.
This largely simplifies the calculations below. 
For example, the stellar photon density at the wind shock region as well as 
the IC scattering angle for the observed gamma-rays 
can be approximated as constant at a given phase. 

The shocked relativistic wind, at the distance of $r_{\rm s}$ from the 
pulsar, is magnetized with  strength of \citep[e.g.,][]{TA97}:
\begin{eqnarray}
\label{eq:mag1}
B & \sim & 3 
\left( \frac{\sigma \dot{E}_{\rm p}}{(1+\sigma) c r_{\rm s}^2} \right)^{1/2} \\
& \simeq & 1.0
\left(\frac{\sigma}{0.01}\right)^{1/2}
\left(\frac{\dot{E}_{\rm p}}{8\times 10^{35}\ \rm erg\ s^{-1}}\right)^{1/2}
\left(\frac{r_{\rm s}}{0.1\rm AU}\right)^{-1} \rm G,
\label{eq:mag2}
\end{eqnarray}
where 
$\dot{E}_{\rm p}$ is the spindown power of the pulsar, and 
$\sigma$ is the magnetization factor defined by the ratio of 
the Poynting and kinetic energy flux in the pulsar wind. 
A factor of 3 in equation~(\ref{eq:mag1}) is introduced to account for 
the shock compression, and 
$\sigma \ll 1$ is assumed 
in obtaining equation~(\ref{eq:mag2}). 
The spindown power is measured to be 
$\dot{E}_{\rm p} \simeq 8\times 10^{35}\ \rm erg\ s^{-1}$.
The magnetization factor was adopted as $\sigma = 0.02$ in \citet{TA97}.
The Kennel \& Coroniti model of the Crab Nebula yielded 
$\sigma = 0.003$ \citep{KC84a}, which was determined to satisfy 
the flow and pressure boundary conditions at the outer edge of the nebula.

The radiation energy density at a distance $d$ 
from the Be star can be estimated as:
\begin{eqnarray}
\label{eq:rad1}
U_{\rm ph} & \sim & 
\frac{L_\star}{4\pi c d^2} \\
& \simeq & 1.2
\left(\frac{L_\star}{10^{38}\ \rm erg\ s^{-1}}\right)
\left(\frac{d}{\rm AU}\right)^{-2} \ \rm erg\ cm^{-3},
\label{eq:rad2}
\end{eqnarray}
where $L_\star$ is the luminosity of the Be star:
$L_\star \simeq 7\times 10^{37}\ \rm erg\ s^{-1}$ for 
$T_{\rm eff} = 27000$ K and $R_\star = 6 R_\sun$.
($L_\star$ is somewhat uncertain. 
For a larger stellar radius of $R_\star = 10 R_\sun$, 
$L_\star$ becomes larger  by a factor of 3.)
The stellar light dominates over the magnetic field of the pulsar wind 
at the termination shock. 
However, because the IC scattering of  TeV electrons on stellar photons 
occurs in the deep Klein-Nishina regime, 
the IC emission can be largely suppressed. 
The synchrotron cooling is expected to overtake 
the IC cooling at high energies. 
The larger luminosity of the observed synchrotron X-ray emission 
compared with the IC gamma-rays during periastron passage 
suggests that the synchrotron cooling is indeed faster than the 
Klein-Nishina IC cooling  at TeV energies. 

\begin{figure}[htbp]
\epsscale{1.0}
\plotone{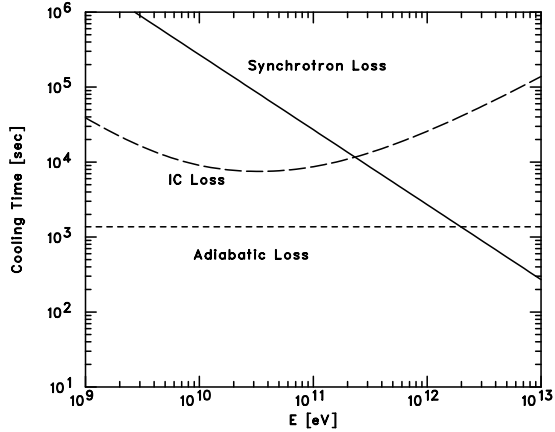}
\caption{\small  
Cooling timescales as a function of electron (positron) energy $E=\gamma m_ec^2$
at $\tau = 30$ days with the following parameters:
$B= 0.4$~G, $d= 1.8$~AU, 
$t_{\rm ad} = 10  r_{\rm s}/c$, and $r_{\rm s} = 0.15d$.
\label{fig:lifetime} }
\end{figure}

Given the postshock flow speed of $\sim c/3$ in the pulsar wind, 
the adiabatic loss can compete with the radiative losses. 
The adiabatic loss rate is adopted as
 $\dot{\gamma}_{\rm ad} = - \gamma / t_{\rm ad}$ with 
 $t_{\rm ad} = \xi  r_{\rm s}/c$. 
 The factor $\xi$, likely $\xi \sim 3\mbox{--}10$, 
 is treated as a free parameter. In Figure~\ref{fig:lifetime}, we show 
the energy-dependence of 
the cooling timescales both for radiative (synchrotron and IC) and 
 non-radiative (adiabatic loss) losses, at $\tau =30$ days, 
 where the importance of adiabatic losses is illustrated.
Recently, it has been shown that 
the hydrodynamics of the interaction of the pulsar and stellar winds 
do predict the predominance of the adiabatic loss over the 
radiative losses \citep{Mitya08}.
The discussion on the cooling regimes in the \PSR\ system 
can be found in \citet{TAK94}.

The Lorentz factor of the relativistic wind is denoted by $\gamma_1$;
the unshocked pulsar wind is presumed to 
contain \pair\  with  an energy of $\gamma_1 m_ec^2$. 
Using $\sigma = 0.003$, 
\citet{KC84b} obtained $\gamma_1 = 10^6$ for the Crab pulsar 
to account for the observed nonthermal radiation from the nebula. 

Unshocked cold \pair\ pairs  in the relativistic wind are 
assumed to be accelerated at the termination shock and 
be injected into the downstream postshock flow.
The energy distribution of the injected nonthermal \pair\ pair is described 
by  $Q(\gamma ) = Q_0 \gamma^{-p} \exp (-\gamma / \gamma_{\rm m})$ for 
$\gamma \geq \gamma_1$. A low-energy cutoff with a step function is assumed:
 $Q(\gamma ) = 0$ for $\gamma <  \gamma_1$. 
Results of simulation by \citet{Hoshino92} for the relativistic collisionless 
shock in electron-positron-ion plasma 
suggest a power-law spectrum with $p \sim 2$ can be formed above
$\gamma_1$, and a fraction 
$\epsilon \sim 0.2$ of the flow energy can be transferred to the 
nonthermal \pair\ pairs, under the condition that 
the upstream flow energy carried by ions exceeds that of pairs. 
We note that $\gamma_{\rm m}$ should be limited by 
rapid cooling of  the \pair\ pairs at  highest energies. 
The OSSE results in the year 1994 
indicated  $\gamma_{\rm m} > 10 \gamma_1$ \citep{Grove95}.
The normalization $Q_0$ is determined by the spindown power of 
the pulsar and the parameter $\epsilon$, through the relation:
\begin{equation}
\label{eq:epsilon}
\epsilon \dot{E}_{\rm p} =  m_ec^2 
\int_{\gamma_1}^{\infty}  \gamma Q(\gamma) d\gamma . 
\end{equation}

The energy distribution of the accelerated \pair\ pairs 
integrated over the entire radiation zone, $N(\gamma )$,  
can be derived by 
the well-known kinetic equation.
We consider simplified cases where 
 both the magnetic field and radiation field do not change 
 within the radiating zone. 
The cooling timescale of the accelerated \pair\ pairs 
is much shorter than the dynamical timescale 
of the binary. Therefore, the energy distribution $N(\phi, \gamma )$ 
at a given orbital phase $\phi$ is reduced to be a steady-state solution
\citep[e.g.,][]{Mitya07}:
\begin{equation}
\label{eq:N}
N(\phi, \gamma) = \frac{1}{b(\phi, \gamma)}
 \int_{\gamma}^{\infty}  Q(\phi, \gamma^{\prime}) d\gamma^{\prime},
\end{equation}
where $b(\phi, \gamma) = - (\dot{\gamma}_{\rm syn}
+\dot{\gamma}_{\rm ic} +\dot{\gamma}_{\rm ad})$ is the 
energy loss rate as given by a sum of the synchrotron, IC, and adiabatic loss rates. 

We are now ready to calculate the synchrotron and IC spectra 
using $N(\phi, \gamma)$, and compare them with the observations. 
Anisotropic IC scattering of stellar photons is taken into account by utilizing 
the IC kernel derived by \citet{AA81}, as 
the IC gamma-ray emission depends largely on the angle between the star-pulsar 
and pulsar-observer directions \citep[][]{KA05}.
We assumed a stellar radiation field 
to be radial rays originating from the point source, the Be star. 
The effect of finite size of the star does not affect the resultant radiation 
spectrum. However, 
 the radiation field 
can be modified appreciably by the presence of the equatorial disk, 
which has to be taken into account by future work to model the 
TeV gamma-ray spectra. 
We calculate the attenuation of TeV gamma-rays 
due to pair production with stellar photons, 
though the $\gamma\gamma$ opacity is small, typically 
$\exp (-\tau_{\gamma\gamma}) \sim 0.9$,  for our choice of the optical luminosity. 
Whereas 
more significant absorption, up to $\exp (-\tau_{\gamma\gamma}) \sim 0.6$, 
can be expected if we adopt  larger stellar luminosity \citep{Dubus06},
this does not affect our main conclusions.

\begin{deluxetable}{rrrrrrrrrr}
\tabletypesize{\small}
\tablecaption{Parameters of Synchrotron-IC models \label{tbl:model}}
\tablewidth{0pt}
\tablehead{
\colhead{} &\colhead{} &  \multicolumn{5}{c}{Fixed} & \colhead{}
& \multicolumn{2}{c}{Free} \\
\cline{3-7} \cline{9-10}\\
\colhead{$\tau$} & \colhead{} & \colhead{$\epsilon$} & \colhead{$\sigma$}
& \colhead{$\gamma_1$} 
& \colhead{$p$} 
& \colhead{$E_{\rm m}$} & \colhead{} 
& \colhead{$\zeta$}
& \colhead{$\xi$}\\
\colhead{} & \colhead{} & \colhead{} & \colhead{}  & \colhead{}  
& \colhead{} & \colhead{(TeV)} & \colhead{} & \colhead{} & \colhead{}
}
\startdata
$-15$ days &  $\cdots$ & 0.1 & 0.01& $4\times10^5$ & 1.9 & 10  & & 0.05 & 3 \\
$+30$ days &  $\cdots$ & 0.1 & 0.01& $4\times 10^5$ & 1.9 & 10  & & 0.15 & 10 \\
$+618$ days &  $\cdots$ & 0.1 & 0.01& $4\times 10^5$ & 1.9 & 10  & & 0.50 & 2
\enddata
\tablecomments{
$\epsilon$ (a fraction of the spin down power 
channelled into the accelerated \pair\ pairs), 
$\sigma$ (the magnetization factor of the pulsar wind), 
$\gamma_1$ (the Lorentz factor of the pulsar wind), 
 $p$ (the acceleration index), 
$E_{\rm m} = \gamma_{\rm m} m_e c^2$ (the maximum energy of 
accelerated pairs),
 $\zeta = r_s/d$ (the distance of the  termination shock from the pulsar 
 divided by the pulsar-Be star separation),  and 
$\xi$ (the parameter to describe the adiabatic loss rate). }
\end{deluxetable}
 
\subsection{Applications to the Observed Spectra}\label{sec:sed1}

To describe the broadband emission of the compactified PWN 
in variable environments, 
two parameters, $\zeta \equiv r_s/d$ and $\xi$, are allowed to vary 
along the orbit.
We note that $\zeta$, which is controlled by ram pressure of stellar wind/disk, 
determines magnetic field strength. 
The two parameters describe physical conditions external to the pulsar wind. 
The fact that we observed a spectral break 
during the disk crossing suggests that the external conditions indeed regulate 
the position of the break. 
The other parameters are 
frozen to be the following values: 
$\epsilon = 0.1$, $\sigma = 0.01$, 
 $\gamma_1 = 4\times  10^5$, 
 $p = 1.9$ and $E_{\rm m} = 10\ \rm TeV$.
They are broadly similar to what have been previously 
obtained in the literature mentioned above, though we made some 
fine-tuning; 
 $\gamma_1$ is adjusted to explain the spectral break, 
and $p$ is chosen to match with the observed X-ray spectral slope 
in Sz7+8 (see below). 
The uncertainties of the fixed parameters are discussed in \S\ref{sec:basic}.
The physical parameters are summarized in Table~\ref{tbl:model}.

\begin{figure*}[htbp]
\epsscale{1.0}
\plotone{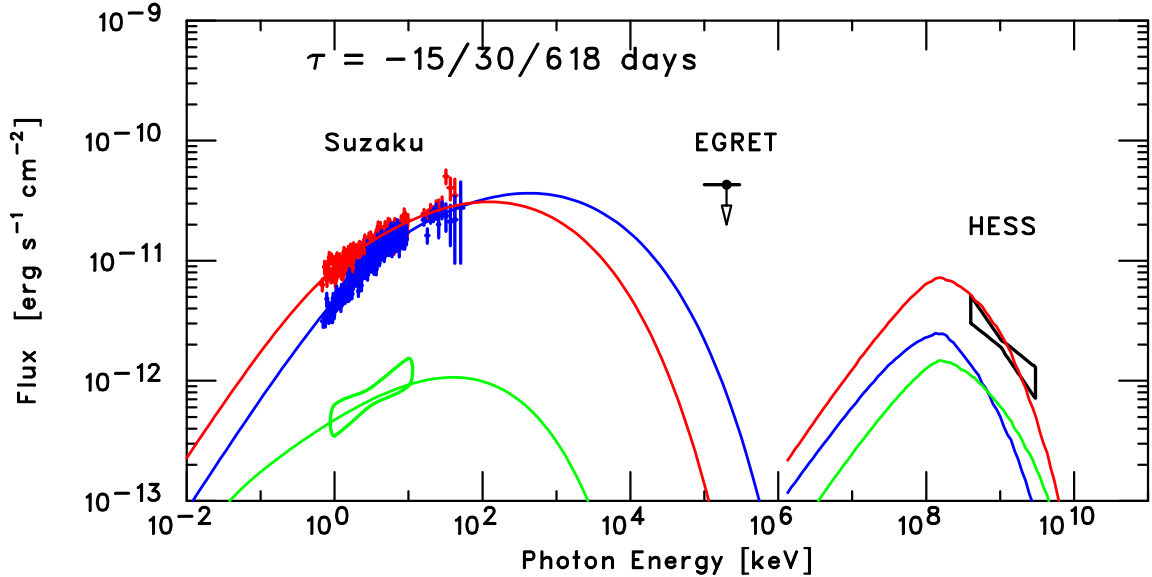}
\caption{\small  
Broadband spectra of the \PSR\ system 
along with the synchrotron (X-ray) and IC (gamma-ray) models.
The average HESS spectrum in 2004 is indicated 
to allow a comparison with the IC calculation. 
The flux upper limit at 200 MeV is obtained with the EGRET observations in 1994.
{\it Red}: Synchrotron-IC model at $\tau = 30$ days in comparison with 
\emph{Suzaku} X-ray spectral data for Sz7+8.
{\it Blue}: $\tau = -15$ days with \emph{Suzaku} Sz3+4 which exhibits a spectral break.
{\it Green}: $\tau = 618$ days (apastron) with 
\emph{ASCA} X-ray spectrum around apastron \citep{Hirayama99}.
\label{fig:sed1} }
\end{figure*}

We first reproduce the Sz7+8 spectrum, which is described by 
a simple power law of $\Gamma = 1.63\pm 0.02$ in a wide 
energy range of 0.6--50 keV, along with the HESS spectrum in 2004 \citep{HESS05}.
  Since our treatment of the radiation field ignores 
 the presence of the equatorial disk, 
 we require only a crude match with the average flux level reported by HESS. 
(The TeV gamma-ray flux changed by a factor of 2--3 from one month to the next.)
Nevertheless, 
the TeV flux and shape put meaningful constraints on the model.
Using the configuration of the binary system at $\tau = +30$~days, 
we found that the following parameters give a reasonable fit to the 
X-ray and gamma-ray spectra:
 $\zeta = 0.15$ ($r_s = 0.27\ \rm AU$ and $B = 0.38\ \rm G$), and $\xi = 10$. 
In Figure~\ref{fig:sed1}, the spectral energy distribution (SED) 
of the synchrotron and IC radiation is drawn. 
The cooling of the accelerated \pair\ responsible for 
the soft X-ray emission is dominated  by adiabatic loss, 
which ensures the hard X-ray spectrum. 
The shape of the TeV gamma-ray spectrum obtained with HESS can be 
reproduced by this modeling as well.

We then proceed to model the Sz3+4 spectrum, 
where a spectral break was observed. 
We  relate the X-ray spectral break 
 to the low-energy cutoff in the electron injection function, 
 which can be ascribed to the Lorentz factor of the pulsar wind  $\gamma_1$.
 In Figure~\ref{fig:sed1}, 
we present calculation of the synchrotron and IC models at $\tau= -15$ days 
with  $\zeta = 0.05$
 ($r_s = 0.057\ \rm AU$ and $B = 1.8\ \rm G$),
 and $\xi = 3$. 
The X-ray spectral shape, the very flat spectrum below 
$\varepsilon_{\rm br}$ in particular, 
 is nicely reproduced by this model. 
With this scenario,  the appearance of the break in the X-ray band is due to 
 the enhanced magnetic field. 
Our model predicts lower flux in the TeV band at the time of 
a higher break energy. 

The observed break energy is related to the Lorentz factor of the 
pulsar wind:
\begin{eqnarray}
\varepsilon_{\rm br} 
& = & \sqrt{\frac{3}{2}} \frac{e\hbar }{m_e c^2} B \gamma_1^2 \\
& \simeq &4 \left(\frac{B}{1.8\ \rm G} \right)
\left(\frac{\gamma_1}{4\times 10^5}\right)^{2} \ \rm keV.
\end{eqnarray}
It is interesting to note that the X-ray spectrum below $\varepsilon_{\rm br}$ 
is predicted to have a universal shape as long as the cooling is 
dominated by adiabatic loss and the energy distribution 
$Q(\gamma )$ below the low-energy cutoff is harder than $\gamma^{-1}$.
Given the adiabatic-loss dominance ($b\propto \gamma$), 
equation~(\ref{eq:N}) yields 
\begin{equation}
N(\gamma ) \propto \gamma^{-1} \ \ \mbox{for} \ \ \gamma \leq \gamma_1,
\end{equation}
irrespective of acceleration index $p$. 
This form of the energy distribution of the cooled pair population 
is translated into  a synchrotron power-law spectrum with 
photon index $\Gamma =1$.
Therefore, the synchrotron spectrum has a universal form 
with $\Gamma =1$ for $\varepsilon \ll \varepsilon_{\rm br}$. 
In fact, the result of the \emph{Suzaku} observations,  namely 
$\Gamma \simeq 1.2$ 
immediately below $\varepsilon_{\rm br}$, agrees well with 
what theoretically expected. 

Let us check if this model can accommodate the X-ray observations 
around apastron, where the physical conditions change dramatically 
from periastron.
We found the following parameters are compatible with the X-ray 
observations at apastron: 
 $\zeta = 0.5$ ($r_s = 4.7\ \rm AU$ and $B = 0.02\ \rm G$), and
 $\xi = 2$. 
 (A condition of  $\zeta \ll 1$ is not well satisfied, 
 so our calculations need a revision if we aim at performing precise modeling. )
In Figure~\ref{fig:sed1}, we show the \emph{ASCA} spectrum and 
the model  appropriate for apastron, $\tau = 618$ day.
 The effective photon index of the model is
 $\Gamma_{\rm syn} = 1.7$, which is in agreement with the 
 X-ray spectrum, 
 $\Gamma = 1.6 \pm 0.2$, obtained with \emph{ASCA} \citep{Hirayama99}.
 
\subsection{Basic Properties of the Pulsar Wind of PSR B1259$-$63}\label{sec:basic}

One of the most fundamental parameters of a pulsar wind, 
the Lorentz factor of the wind $\gamma_1$, can be estimated by our modeling 
in the previous section 
through 
the identification of the spectral break with the low-energy cutoff of the 
accelerated pairs. 
We adopted $\gamma_1 = 4\times 10^5$ to account for the observed break 
position. 
 It  depends only on 
the square root of postshock magnetic field, $\propto \sqrt{B}$, 
which can be constrained well by the X-ray to TeV gamma-ray flux ratio.
Therefore, $\gamma_1$ is hard to be altered more than by a factor of few. 

It should be noted that the wind Lorentz factor will  possibly be measured 
by the \emph{Fermi} Gamma-ray Space Telescope during the next 
periastron passage. The Comptonization of the {\it unshocked} pulsar wind 
inevitably  leads to the formation of an additional bump-like gamma-ray spectrum 
at $\varepsilon \sim \gamma_1 m_e c^2$ (assuming the Klein-Nishina limit), 
which overwhelms the IC component 
 in Fig.~\ref{fig:sed1} in the 0.1--10 GeV band
\citep{Mitya07}.
The X-ray and  GeV-to-TeV gamma-ray data during the next 
periastron passage will be crucial to test the models of  the \PSR\ system, 
and the long-standing pulsar wind paradigm \citep{KC84a} 
in general. 

The other fundamental parameter in the standard pulsar wind paradigm 
is the magnetization factor $\sigma$.
While the magnetic field strength can be well constrained by 
the X-ray and TeV flux ratio as $B\sim 0.5$~G during  the disk  crossings, 
the $\sigma$-parameter itself is allowed to vary in a wide range of 
$10^{-4} < \sigma < 1$ depending on $r_s$.
The condition $r_s < d$ required to explain the spectral variability 
corresponds to $\sigma < 1$, and 
the energetics requirement of $\epsilon < 1$ (together with $\xi \sim 10$) 
gives a lower bound as $\sigma > 10^{-4}$. 

The magnetic field of $B\sim 0.5$~G constrains 
particle acceleration mechanism(s) at work in the termination shock region. 
Synchrotron cooling time of 10 TeV electrons is about 100 sec. 
To overcome the fast cooling, the acceleration timescale of 
$t_{\rm acc} \sim 100\ {\rm sec} \sim 50 r_{\rm L}/c$
 is necessary at $E=10$ TeV, where $r_{\rm L}$ denotes 
the Larmor radius of pairs  \citep[see][]{Mitya07}.
According to the magnetosonic acceleration model of \citet{Hoshino92} 
where the magnetosonic waves collectively emitted by the shock-reflected 
protons (or heavy ions) are resonantly absorbed by pairs,
the inverse of the gyrofrequency of protons  with 
a Lorentz factor of $\gamma_1$ determines the acceleration timescale; 
interestingly, $t_{\rm acc} \sim m_{\rm p} c\gamma_1 /(eB) \sim 50$ sec 
is compatible with the requirement posed by our modeling. 

The efficiency with which the pulsar spindown power 
is transferred into the nonthermal \pair pairs, 
namely $\epsilon$ that is defined by equation (\ref{eq:epsilon}),
was adopted as $\epsilon = 0.1$ in our modeling,
which is also consistent with the numerical result of \citet{Hoshino92}.
It should be noted, however, that the determination of $\epsilon$ would be 
subject to the degree of   relativistic Doppler boosting if 
possible relativistic motion of the postshock flow \citep{Bogovalov08} 
plays a major role in the nonthermal emission.

\subsection{Alternative Model for the Spectral Break} 

We have explored another possibility to explain the X-ray spectral 
break---the observed break may be accounted for by the transition of main 
cooling mechanisms; 
below the break, cooling of pairs is assumed to be 
 IC cooling in the Klein-Nishina regime, while 
synchrotron cooling takes over above the break. 
Due to the inefficient IC cooling in the Klein-Nishina, the electron 
energy distribution can be much harder in a certain energy band, 
as compared to the energy bands wherein 
the IC cooling in the Thomson regime or synchrotron cooling are dominant.
With this interpretation, 
$\xi \ga 100$ is necessary to make the adiabatic loss unimportant. 

To match the flat X-ray spectrum below $\varepsilon_{\rm br}$, 
the acceleration index should be very flat, $p \simeq 1.0$.
However, the very flat spectrum of accelerated \pair\ 
would be inconsistent with the conventional particle acceleration model 
as well as the observations of PWNe on parsec scales. 
Also, the significant reduction of the X-ray flux around apastron 
requires a drastic change of $\xi$ (say, from $\xi = 100$ to $\xi = 3$). 
In view of these uncomfortable 
requirements, we prefer the adiabatic-loss-dominated 
model in \S\ref{sec:sed1}, where the X-ray break is ascribed to the 
appearance of the Lorentz factor of the relativistic pulsar wind.

\section{Conclusions}

We have presented results of  new X-ray observations of the 
pulsar-Be star binary system \PSR\ performed eight times near 
periastron using the \emph{Suzaku} satellite. 
The monitoring observations with \emph{Suzaku}, coupled with previous data, 
allow us to identify the two bump in the X-ray lightcurve around periastron, 
which can be ascribed to the pulsar's entrance to the equatorial disk of the Be star. 
In most cases, the X-ray spectra can be well fit by a simple power law, 
as observed in previous observations. 
However, we found evidence of a spectral break when the X-ray spectral shape 
below 10 keV is described formally by a rather hard power law 
with photon index $\Gamma \sim 1.3$. 
Applying a broken power-law model to the Sz3+4 spectrum 
in the 0.6--50 keV band, 
we found that the index changes 
by $\Delta \Gamma \simeq 0.4$, from  
 $\Gamma_1 = 1.25^{+0.02}_{-0.04}$ to 
  $\Gamma_2 = 1.66^{+0.05}_{-0.04}$,  
around $\varepsilon_{\rm br} = 4.5^{+0.5}_{-0.2}$ keV.

We considered a compactified PWN model, in which 
the X-ray emission is accounted for by shock-accelerated relativistic \pair\ pairs 
via synchrotron process. 
The spectral properties of the X-ray and TeV gamma-ray emission 
are found to be reconcilable in a compactified PWN 
model whose parameters are adjusted within a reasonable range. 
We ascribed the spectral break to the low-energy cutoff of the accelerated 
\pair\ pairs, which is supposed to be linked directly to the wind Lorentz factor. 
We here emphasize that our choice of parameters (see Table \ref{tbl:model}) 
is definitely not the unique one that can fit the observed spectra. 
Nevertheless, the  value of the wind Lorentz factor, 
$\gamma_1 \sim 4\times 10^5$, determined  by the observed break 
energy in the X-ray band 
is robust within a factor of few, because it depends only on 
the square root of postshock magnetic field, 
which in turn can be deduced from the X-ray to TeV gamma-ray flux ratio 
as $B=0.4\mbox{--}2$~G during the periastron passage. 
Also, the magnetic field estimate indicates the magnetization factor 
should be  in a  range of $10^{-4} < \sigma < 1$.
The determination of the wind Lorentz factor through the X-ray spectral break, 
if confirmed, provides a direct probe of the standard pulsar wind paradigm 
in which  pulsar's spindown power is supposed to be 
carried off by a relativistic wind with Lorentz factor $\gamma_1 > 10^4$.

\acknowledgments

We acknowledge the useful suggestions of the anonymous referee, which 
improved the manuscript. 
We wish to thank Felix Aharonian and Masha Chernyakova for fruitful discussions. 
T. Tanaka is supported by research fellowships of the 
Japan Society for the Promotion of Science for Young Scientists.

{\it Facilities:} \facility{Suzaku ()}






\end{document}